# Speech prosody and remote experiments: a technical report

*Giuseppe Magistro*[1][2]

[1]Department of linguistics, University of Ghent, Belgium
[2]Department of linguistics, University of Konstanz, Germany

giuseppe.magistro@ugent.be, giuseppe.magistro@uni-konstanz.de

## Abstract

The aim of this paper is twofold. First, we present a review of different recording options for gathering prosodic data in the event that fieldwork is impracticable (e.g. due to pandemics). Under this light, we mimic a long-distance reading task experiment using different software and hardware synchronously. In order to evaluate the employed methodologies, we extract noise levels and frequency manipulation of the recordings. Subsequently, we examine the impact of the different recordings onto linguistic variables, such as the pitch curves and values. We also include a discussion on experimental practicalities. After balancing these factors, we decree an online platform, Zencastr, as the most affordable and practical for acoustic data collection. Secondly, we want to open up a debate on the most optimal remote methodology that researchers on speech prosody can deploy.
.

## 1. Introduction

Interactive fieldwork sessions for gathering human speech data have become the standard procedure in phonetics and phonology [1]. However, after the spread of the SARS-CoV-2, such expeditions were hindered, making it impossible to gather good quality data without required instrumentation [2]. Since the virus outburst, many attempts to recover acoustic data were made. However, comparability issues arose inasmuch as disparate recording hardware and software have been used to compensate such limitations. This variability of instrumentation could be accounted for resorting to mixed modeling [2], but this yields further intricacy and possibly leads to statistical overfitting in the model. From a more practical perspective, the feasibility of the experiment may be endangered without the supervision of researchers on the spot. In order to review these issues and find an optimal solution, we ran an experiment simulating different long-distance recording sessions synchronously.

## 2. Methodology

### 2.1. Rationale of the experiment

Before presenting the experiment, it is worth premising that the following experimentation and discussion are directly applicable in research on speech prosody. Moreover, we simulated recording sessions assuming that the target community of informants can easily access electronic devices and have minimal digital literacy. We do not exclude that other peculiar situations and domains of inquiry might require further *ad hoc* adjustments.

### 2.2. Experimental setting

We recorded a 25 year old male speaker from Italy (Bari), reading a list of words and sentences in Italian. These recordings were simultaneously made by three devices. First, we took as reference device a headset microphone Røde HS1 with Fuzzy Windbuster, plugged in a solid-state recorder Marantz PMD661MKII. We took this microphone as a control standard since similar models are commonly used in linguistic documentation. On the left side of the speaker, at 20 cm at 330°, an Android tablet was placed (Samsung Galaxy Tab S6) and the recording was made using its inbuilt MEMS microphone and an application for recording (VoiceRecorder [3] ). At 20 cm but at 30° degrees an Android phone was placed recording with the same application (Xiaomi Redmi Note 6 PRO). In front of the speaker, at 20 cm we placed a laptop (Dell Latitude 5501) and recorded with its inbuilt microphone using both SpeechRecorder software [4] and an online platform designed for broadcasting, Zencastr. [5]. As well as the equidistance of device placement, we made sure that the polarity of the microphones was omnidirectional since they were placed at different axes from speakers (see Figure 1). Recordings were taken in a normal room, to simulate remote recording conditions. The five simultaneous recordings were sampled at 44.1 kHz and saved in .wav format.

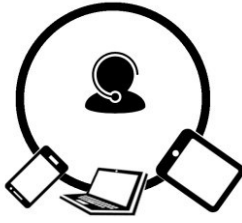

Figure 1: *Placement of the devices in the simultaneous recording experiment*

**2.3. Experimental design**

The stimuli contained a set of 15 isolated words containing stressed /i/, /a/, /u/ (as these are shown to suffer from poor quality recordings as seen in [6]), a set of 15 declarative sentences with the aforementioned words in nuclear position, and a set of 15 interrogative sentences with the same segmental layout.

Table 1: *Segmental layout of the target words*

| mira | milo | mito |
| mare | male | mate |
| muro | mulo | muto |

**2.4. Data preparation, first inspection and analysis**

Recordings were cut to correct delays in starting and finishing the session on each device. As preliminary analysis, we checked the quality of recording, in terms of performance and fidelity. In order to see whether different noise levels are found across devices, we extracted with Praat [7] the respective Harmonicsto-Noise Ratio (henceforth HNR), the ratio of energy of periodic signal and noise (formula 1).

$$HNR = 10\log_{10}\left(\tfrac{P}{N}\right) \quad (1)$$

P stands for the percentage of energy in the periodic signal and N in noise. To further evaluate the fidelity to frequency across the devices, long-term spectra were analysed in Audacity [8]. In order to quantify the spectral manipulation occurring in the frequency domain of interest (0-500 Hz), the slope of an interpolated line was computed.

After evaluating the general performance of the recordings, we analysed the linguistic variables concerned. First, pitch curves for the target sentences were plotted using the autocorrelation algorithm in Praat and visually inspected. After observation of pitch curves, we tested whether the curves for an interrogative sentence differed sensitively depending on the type of recording. Since the pitch curves in question have a non-linear distribution, we fitted a Generalized Additive Mixed Model (GAMM), [9, 10] using the *mgcv* package [11] and the itsadug package [12] in R. Ultimately, pitch values on the stressed syllables of the target words were computed in the vocalic onset and compared with statistical independence tests.

In order to choose the most appropriate recording strategy, we combined these results with considerations about the ease and feasibility of such remote experiment.

## 3. Results and considerations

### 3.1. Results

The extraction of HNR values confirmed that the Røde microphone sets the standard, as it has the highest median value in the comparison (Table 1).

Table 1: *Median HNR values.*

| Recording | Median HNR |
| --- | --- |
| Røde | 13.86 |
| Tablet | 11.95 |
| Phone | 10.84 |
| Zencastr | 10.27 |
| SpeechRecorder | 10.18 |

Lower HNRs may impact on the auto-correlation algorithm, generating pitch miscalculation and then forcing researcher to set the pitch window manually in the analysis. [13].

With regard to spectral balance, we report in table 2 the coefficients of the interpolated line in the frequency bands 80300 Hz.

Table 2: *Spectral tilt at 80-300 Hz*

| Recording | Spectral tilt |
| --- | --- |
| Røde | 0.013 |
| Tablet | 0.045 |
| Phone | 0.059 |
| Zencastr | 0.018 |
| SpeechRecorder | 0.022 |

Røde displayed a flat frequency response (an ideal balanced coefficient would be 0). The recordings of Speechrecorder and Zencastr are very similar in this regard, and performed best with the smallest coefficients. The phone recording had a very high coefficient, since lower frequencies were massively attenuated, resulting in an extremely unbalanced long-term spectrum. After checking general aspects of recording quality, we moved on to describing how linguistic variables were affected by the type of recording. First, we compared the pitch curves of the sentences qualitatively. As for reference, we present the pitch curves for the interrogative sentence "*hai preso la mira?*" (have you taken aim?). As already documented in previous studies on the speaker's variety [14, 15], information-seeking yes-no questions display a L+H* nuclear accent with a L-L% phrasal tones (using the ToBI transcription system, [15]). Such pattern can be uniformly appreciated in the recordings. However, when comparing the prosodic details, artefacts were spotted. In figure 2, we plotted the pitch curves of the interrogative sentence for two different recordings in the same time window.

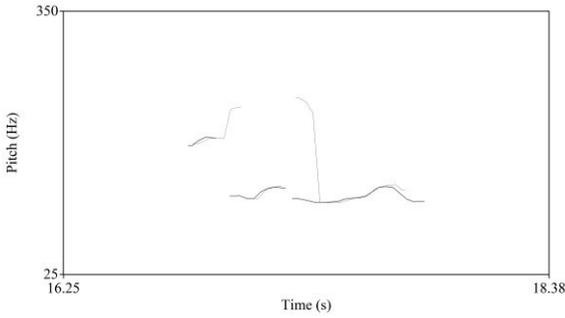

*Figure 2: Pitch curves: the thick line represents the recording by the standard Røde, the thin line represents SpeechRecorder.*

As suspected, poor HNR levels in SpeechRecorder caused unwanted results in the auto-correlation algorithm, such as octave jumps and pitch omissions. Similar inaccuracies were also detected in the phone recording. This leakage can be manually circumvented by the researchers, but similar solutions can be inconvenient and may harm the analysis workflow. More importantly, as shown by Figure 3, time-domain features can be vastly comprised. Although the general trend of the pitch curve is maintained, the time dimension is mismatched, provoking a linguistic feature such as alignment [16] to be in jeopardy.

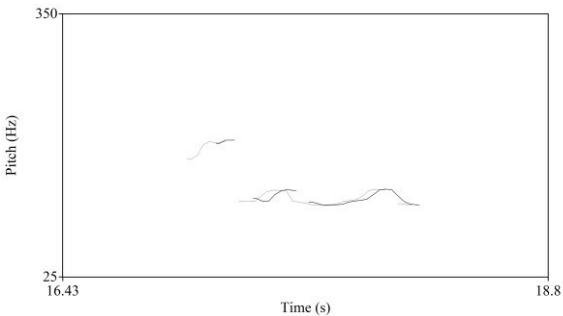

*Figure 3: Pitch curves: the thick line represents the recording by the standard Røde, the thin line represents the tablet.*

Overall, the comparison between *Røde* and *Zencastr* showed the least deviant results, excepting some residual timing problems that still surface (Fig.4).

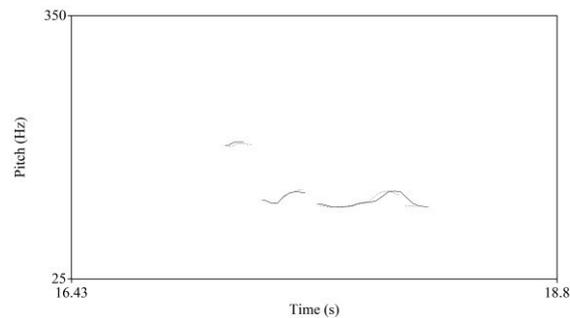

*Figure 4: Pitch curves: the thick line represents the recording by the standard Røde, the thin line represents the recording using Zencastr.*

Quantitatively, after applying manual correction of octave jumps, Generalized Addictive Mixed Models showed no significant statistical difference among all curves ($R^2$= 0.753) (table 3) . In figure 5, we also report the prediction plot for the curves analysed.

Table 3: *GAMM results for the predictor recording strategy*

| Recording | Estim. coeff. | Std.error | p-value |
|---|---|---|---|
| Intercept (Røde) | 137.568 | 5.271 | *** |
| Speechrecorder | 11.353 | 7.454 7.454 | .136 |
| Tablet | -2.561 | | .733 |
| Phone | -2.022 | 7.454 | .788 |
| Zencastr | -2.500 | 7.454 | .739 |

Approximate significance of smooth terms (time instant of measurement)

| | df | F | p-value |
|---|---|---|---|
| **s(time)** | 7.74 | 17.16 | *** |

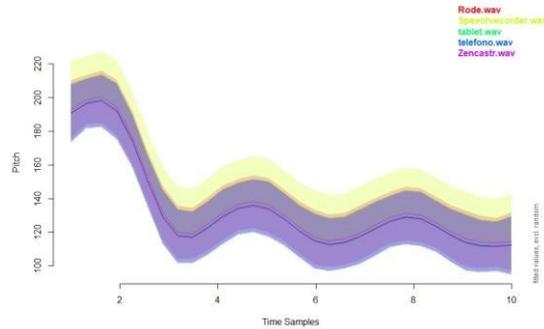

*Figure 5: Prediction plot for the type of recording with the pitch curve of the analysed interrogative sentence.*

These results must be interpreted in light of the nature of the modelling, which fits few samples at the same time in nonlinear curves. Although this result shows that the general pitch tendency is reproduced evenly throughout the corpus, timing details might still be at stake.

Finally, we compared mean absolute pitch values on the vowel segments contained in the target words of questions (N = 45). After looking at the normality of distribution and homogeneity of variance, a Kruskal-Wallis H Test was conducted to examine the differences in pitch values in variation with the types of recording. No significant differences ($\chi^2$= 4.58, p = .47, df = 5) were found among the five recordings. Similar low error rates due to the device bias on pitch have been already reported in [17, 18, 19]. Despite the fact that f0 values seem robust, errors in pitch visualization may arise, as shown by the previous discussion.

These results shed optimism on the evaluation of pitch curves, but also cast shadows on possible errors that have to be corrected, which might be an inconvenient practice for researchers.

### 3.2. Considerations on the feasibility of experiment

Although qualitative considerations on the feasibility of remote experiments are rather impressionistic, they must be taken under scrutiny.

First of all, sending professional equipment to the speaker might result cumbersome and endanger the device. Second, the device designated as standard requires installation, preliminary tests, and appropriate maintenance [20]. Consequences of poorly-tested settings showed up in the mock experiment (gain levels and therefore clipping). Android recording applications also need to be installed, but they are more intuitive and do not require technical training. The PC software SpeechRecorder might result convoluted, as it requires uploading an XML file with the stimuli to be read.

Another hurdle is starting the recording: this appeared to be inconvenient for the special case of SpeechRecorder. When using this software, speakers must remember to start the recording for every stimulus and check that they are not running out of the time at their disposal, since the XML script sets a fixed amount of time per recording window. Similar problems appear with other mobile apps but in a less problematic way. Contrarily, linguists can start the recording themselves in Zencastr and speakers are only required to access the invitation URL via browser and give permission for the activation of the laptop microphone.

When it comes to stimuli prompting, SpeechRecorder is the gold standard, as it shows the sentences to be read full-screen and does not involve further windows or external material. Some Android apps allow recording in the background while reading a text file, but this requires speakers' dexterity in multitasking. On the other hand, each Zencastr session comes with an inbuilt chat, that, albeit small, would allow researcher to share (only) textual prompts.

An additional point is retrieving the recording. Almost every tool compels speakers to find files in their folders and share them with the researchers. Such a final step would inevitably lead to further complexity, as most sharing platforms and email services limit the amount of transferable data. Such a problem can be avoided by sending back the solid state recorder to the researcher. Otherwise, in Zencastr, the users interact online but the recordings are being taken in stereo channels on their local laptops. After finishing the session, the .wav files are uploaded on the platform and can be downloaded directly therefrom. This is an interesting asset, as connectivity issues do not affect the recording. In addition, although the recording session is being held online, there is no quality deterioration due to VoIP protocols [21].

One big limitation of Zencastr is that it is not supported on mobile devices and requires speakers to possess a personal computer. More recently, some crowd-sourcing apps like [22][2] send files to servers via FTP protocol that can be retrieved more easily.

## 4. Discussion

The analysis conducted here is far from exhaustive. Depending on the variables being debated, one could possibly search and find better reviewing methods and solutions. Here, the main aim was to set a discussion on alternative methods for gathering data usable in the intonational phonology framework. Under this light, Zencastr was designated as the most optimal tool in case professional equipment could not be used. Although Zencastr turned out to be more sensitive to noise, pitch extraction produced satisfactory output, both on the frequency and time domain. Overall, pitch was not affected significantly by the type of recording, although some errors were found. These errors might slower and endanger the analysis workflow. Under this light, Zencastr performed better, since it shows less difference than the others in the alignment and pitch tracking compared to the standard-quality recording. Moreover, frequency response was found to be flat in the lower frequency bandwidth, promising to be a good device for other frequency values that might be investigated (e.g., first formant).

With regard to the problem of low HNR levels, we argue that in these cases, researchers can cooperate with the remote speaker to soundproof the room, by favouring noise absorption (using rugs, tapestries, blanket, and other soft material) and blocking other sound transmissions (airways, electronic devices). After some pilot recordings, noise levels should be improved, without altering the spectrum sensitively as other recorders do. Even though these programs do not seem to change the spectrum, researchers should be wary that different hardware is being used by different speakers, resulting in comparability issues. On the other hand, one major asset of Zencastr is its ease of usage, which allows to troubleshoot logistic problems of a long-distance experiment.

Of course, more work and discussion are due, depending on the target language community, the nature of experimentation, and the acoustic details that are being surveyed. Hence, with this work, we also aim to bridge an interdisciplinary connection between experts on telecommunication and linguists, so to find and discuss better options for similar scenarios.